# 'Binge' drinking in the UK: a social network phenomenon

## Paul Ormerod[1] and Greg Wiltshire[2]

### June 2008

1. Volterra Consulting UK and Institute of Advanced Study, University of Durham; corresponding author; pormerod@volterra.co.uk
2. Volterra Consulting UK gwiltshire@volterra.co.uk

*Abstract*

*In this paper, we analyse the recent growth of 'binge' drinking in the UK. This means the rapid consumption of large amounts of alcohol, especially by young people, leading to serious anti-social and criminal behaviour in urban centres. This phenomenon has grown very rapidly.*

*British soccer fans have often exhibited this kind of behaviour abroad, but it has become widespread amongst young people within Britain itself. Vomiting, collapsing in the street, shouting and chanting loudly, intimidating passers-by and fighting are now regular night-time features of many British towns and cities. A particularly disturbing aspect is the huge rise in drunken and anti-social behaviour amongst young females.*

*Increasingly, policy makers in the West are concerned about how not just to regulate but to alter social behaviour. Smoking and obesity are obvious examples, and in the UK 'binge' drinking has become a focus of acute policy concern.*

*We show how a simple agent based model approach, combined with a limited amount of easily acquired information, can provide useful insights for policy makers in the context of behavioural regulation.*

*We show that the hypothesis that the rise in binge drinking is a fashion-related phenomenon, with imitative behaviour spreading across social networks, is sufficient to account for the empirically observed patterns of binge drinking behaviour.*

*The results show that a small world network, rather than a scale-free or random one, offers the best description of the data.*

## 1. Introduction

In this paper, we analyse the recent growth of 'binge' drinking in the UK. By this, we mean the rapid consumption of large amounts of alcohol, especially by young people, leading to anti-social behaviour in urban centres. British soccer fans have often exhibited this kind of behaviour abroad, but it has become widespread amongst young people within Britain itself. Vomiting, collapsing in the street, shouting and chanting loudly, intimidating passers-by and fighting are now regular night-time features of many British towns and cities. A particularly disturbing aspect is the huge rise in drunken and anti-social behaviour amongst young females.

The phenomenon is of serious concern to the British government, not merely for the anti-social behaviour related to it, but because of the longer term health implications for young people of massive intakes of alcohol in very short periods of time.

There is a growing literature which demonstrates the importance of social networks for consumer choice in what might be termed 'regular' consumer markets. A popular reference, for example, on this is [1]. The concept of the 'tipping point' is used to explain on why some books, films and music emerge out of obscurity with small marketing budgets to become popular hits when many *a priori* indistinguishable efforts fail to rise above the noise. A much more formal analysis of the importance of social networks in determining success or failure in the film industry is [2].

In many social and economic contexts, individuals are faced with a choice between two alternative actions, and their decision depends, at least in part, on the actions of other individuals. Ref [3] describes this class of problem as one of 'binary decisions with externalities'. An important feature of such systems is that they are 'robust yet fragile' [4,5]. In other words, behaviour may remain stable for long periods of time and then suddenly exhibit a cascade in which behaviour changes on a large scale across the individual within the system.

Two recent American studies [6,7] using the Framingham Heart Study data base [8] have demonstrated the importance of social networks in determining the behaviour of individuals on matters of public health, specifically obesity and smoking. The Framingham data base contains detailed information on over 12,000 individuals, monitored over more than three decades since 1971.

The social networks of individuals on this data base have been important determinants of both the spread of obesity and the reduction in smoking over this period. In terms of obesity, for example, the chance of any individual being obese increased by 57 per cent if he or she had a friend who became obese. When a spouse stopped smoking, the other was 67 per cent less likely to smoke.

The aim of this paper is to examine the extent to which the sudden emergence of the binge drinking problem in the UK can be explained as a social network phenomenon. We use the methodology developed in [9]. An agent based model is set up in which agents face the binary decision on whether or not to binge drink. Transmission of binge drinking behaviour across agents connected on a social network is determined according to a threshold rule. The theoretical model is calibrated against empirical evidence. The approach described here can be used more generally in areas where policy makers are interested in regulating and altering agent behaviour.

Section 2 describes the basic data, Section 3 the initial evidence for the existence of imitative behaviour on social networks, and section 4 the theoretical model and results.

## 2.    The data

In this particular context, no longitudinal survey such as the Framingham Heart Study exists. So data was gathered using standard survey techniques. The market research company FDS interviewed 504 18-24 year-olds in the UK using an online survey based on MyVoice Panel. Of the respondents, 258 (51 per cent) were male and 246 (49 per cent) were female. The sample group was selected to reflect a demographic which is believed to represent a particular problem in terms of alcohol consumption.

Definitions of heavy drinking vary widely [10] and changes to the standard definitions can have a significant impact on the reported incidence of alcohol misuse. For example, the latest Office for National Statistics report on alcohol consumption in the UK [11] introduced a revised methodology for estimating the proportions of heavy drinkers within the population, taking into account increased alcohol strengths and larger drink sizes. This results in increased counts of heavy drinkers in all age and gender categories, even though the underlying data have not changed. For people aged 16-24, for example, the proportion of women identified as heavy drinkers rises from 29 per cent to 40 per cent.

The focus of this study is not on heavy drinking as such, but on drinking behaviour which is likely to lead to anti-social behaviour i.e. binge drinking.

An individual might regularly drink a fairly large quantity of alcohol but (being habituated) might not subjectively experience this as 'bingeing', i.e. might not actually feel that they are particularly drunk. Thus, in order to distinguish between regular *binge* drinkers and those who are simply regular *big* drinkers, our definition is based upon a combination of consumption of alcohol (anyone drinking more than 10 drinks in a single session is considered to be drinking enough to get very drunk, regardless of their own perception), and subjective perception – those who at least once a week drink an amount that they had previously specified as being, for them, "enough to get very drunk".

We have therefore defined 'binge drinking' as follows:

> For men, getting drunk on 4 or more drinks OR having 10 or more drinks (but not necessarily getting drunk) at least once a week and for women, getting drunk on 3 or more drinks OR having 10 or more drinks (but not necessarily getting drunk) at least once a week.

This definition therefore captures behaviour that is directed at purposefully getting drunk, and also includes those who drink excessively (i.e. ten or more drinks in a single session) even if the excessive drinking does not cause the drinker to feel drunk.

Overall, 16.2 per cent of respondents qualified as binge drinkers using the definition described above. Of this group, the vast majority reported anti-social behaviour as a result of binge drinking such as shouting or vomiting in the street, getting into a fight.

Scaling up the survey, the figures indicate there are around 950,000 binge drinkers in the UK 18-24 year old population, participating in 1.5 million binge drinking 'events' each week.

## 3. Initial evidence for interaction on social networks as a factor in binge drinking

We analysed the patterns of social interaction for those classified as binge drinkers and compared them to the patterns of non-binge drinkers. We looked at three types of social group which might have an influence on a person's drinking behaviour:

- Family
- Work colleagues
- Friends

Everyone in the survey was asked what they thought about the binge drinking behaviour of people in their social groups. Table 1 shows the results for family members.

*Table 1:     Proportion of family thought to be binge drinkers: for binge drinkers and non-binge drinkers*

| Proportion of family thought to be binge drinkers | Proportion (%) for binge drinkers | Proportion (%) for non-binge drinkers |
|---|---|---|
| All of them | 9 | 3 |
| Almost all of them | 9 | 3 |
| Most of them | 11 | 8 |
| Some of them | 32 | 23 |
| Hardly any of them | 28 | 29 |
| None of them | 10 | 16 |

So, for example, amongst people who binge drink themselves, 18 per cent think that 'all' or 'almost all' their family members also binge drink. This compares to non-binge drinkers, 6 per cent of which think 'all' or 'almost all' their family members binge drink.

There are differences in the perceived behaviour of the family members of binge and non-binge drinkers, although the differences are not dramatic.

These differences are considerably more marked when the behaviour of work colleagues is examined.

*Table 2:  Proportion of colleagues thought to be binge drinkers: for binge drinkers and non-binge drinkers*

| Proportion of colleagues thought to be binge drinkers | Proportion (%) for binge drinkers | Proportion (%) for non-binge drinkers |
|---|---|---|
| All of them | 11 | 1 |
| Almost all of them | 19 | 7 |
| Most of them | 27 | 14 |
| Some of them | 21 | 27 |
| Hardly any of them | 6 | 12 |
| None of them | 5 | 4 |

Here, for example, no less than 30 per cent of binge drinkers think that 'all' or 'almost all' of their work colleagues binge drink, compared to only 8 per cent of non-binge drinkers.

But by far the most dramatic difference is seen in the behaviour of friends[1].

*Table 3:  Proportion of friends thought to be binge drinkers: for binge drinkers and non-binge drinkers*

| Proportion of friends thought to be binge drinkers | Proportion (%) for binge drinkers | Proportion (%) for non-binge drinkers |
|---|---|---|
| **All of them** | **24** | **5** |
| **Almost all of them** | **30** | **10** |
| **Most of them** | **31** | **21** |
| **Some of them** | **12** | **31** |
| **Hardly any of them** | **1** | **13** |
| **None of them** | **2** | **6** |

Table 3 shows that 54 per cent of binge drinkers think that all or almost all of their friends are binge drinkers, compared to 15 per cent of non-binge drinkers for whom all or almost all friends are binge drinkers. Conversely, only 3 per cent of binge drinkers have no or hardly any friends that binge drink, compared to 19 per cent of non-binge drinkers

---

[1] This is confirmed in formal analysis by calculating both the Manhattan and Euclidean norms between the two columns

### 3. The theoretical model and its calibration

Our aim is to establish whether social network effects are a *sufficient* condition to account for the observed binge drinking behaviour in the UK. We know from Tables 1-3 above, especially Table 3 which is now our specific focus, that binge drinkers have different sets of social networks to non-binge drinkers.

We set up a simple agent based model, in which the decision of an agent to become a binge drinker is determined solely by the proportion of friends on his/her network who already are binge drinkers. The paper follows the methodology described in [9], where the issue analysed was whether people on benefits had bank accounts.

We connect agents on different types of social network, specifically a random, a small world and a scale-free network, using both replacement and additional re-wiring in the latter.

Initially, all agents in the model are in state 0, i.e. they are not binge-drinkers. A small percentage (2 per cent) of the total is selected at random to become binge drinkers (state 1).

Each agent is allocated a threshold above which he or she will convert from state 0 to state 1. This is drawn at random from a uniform distribution on [0,U1], where U1 is a variable of the model. The threshold is the proportion of friends who in state 1, above which the agent will switch from state 0 to state 1, otherwise stay in state 0.

We monitor the percolation of state 1 behaviour across the network, and halt the solution when the proportion reaches 16.2 per cent, the estimated number of state 1 agents from the empirical data.

We then examine the networks of the friends of agents in both state 0 and state 1, and to see how closely they correspond to the observed structure set out in Table 3 above. More precisely, we simplify Table 2 slightly, and merge the categories 'all' and 'almost all' into a single one, and do the same for 'none' and 'hardly any'.

A final assumption needs to be made as to what the categories 'all/almost all', 'most', 'some' and 'hardly any/none' mean in percentages. We use the following:

*Table 4: Assigned values for the questionnaire responses.*

| Questionnaire Response | Assigned Corresponding Value | Value For Quartile Denoted As |
|---|---|---|
| 'Hardly any' and 'None' | ≥ 0 and ≤ 25% | $Q_1$ |
| 'Some' | >25% and ≤ 50% | $Q_2$ |
| 'Most' | >50% and ≤ 75% | $Q_3$ |
| 'All' and 'Almost all' | >75% and ≤ 100% | $Q_4$ |

We conducted extensive searches for the best combination(s) of relevant parameters in each of the three types of networks examined.

The initial sweeping of the combinations of model parameters was performed 40,000 times which equated to averaging each parameterisation over 300-1000 runs. The candidate models taken forward from this sifting were then run an additional 1000 times. The range of parameters examined is set out in Table 5.

*Table 5: Parameters used in the generation of the three types of networks*

| Application | Parameter | Description | Value/ Range |
|---|---|---|---|
| General parameters | $n$ | Number of agents in network. | 1000 |
| | L1 | The lower limit to the distribution for the threshold of agents to switch from not binge drinking to binge drinking based on an evaluation of agents connected to them by their social network. | 0 |
| | U1 | The upper limit to the distribution for the threshold of agents to switch from not binge drinking to binge drinking based on an evaluation of agents connected to them by their social network. | 0.4-0.8 |
| | L2 | The lower limit to the distribution for the threshold of agents to switch from binge drinking to not binge drinking based on an evaluation of agents connected to them by their social network. | 1.2 |
| | U2 | The upper limit to the distribution for the threshold of agents to switch from binge drinking to not binge drinking based on an evaluation of agents connected to them by their social network. | 1.2 |
| Small world network | $k$ | Number of adjacent agents each agent is linked to on either side. | 2-10 |
| | $\phi$ | Probability of rewiring a link (either additionally or replacement) when generating network. | 0-0.1 |
| Scale free network | $q$ | Average number of links each agent makes when it is added to network. | 0.5-2 |
| | $\alpha$ | Number of initially completely connected agents before generating network. | 2-8 |
| Random network | $p$ | Probability that any two agents are connected. | 0.002-0.025 |

In order to compare the results for the various models they were scored using the following equation.

$$S = |q_1 - Q_1| + |q_2 - Q_2| + |q_3 - Q_3| + |q_4 - Q_4|$$

where $q_i$ is the percentage in the relevant quartile of the basic data on friends described in Table 3 above, and $Q_i$ is the model-generated proportion when overall 16.2 per cent of agents are in state 1.

Models with a lower *S,* or score value, will more closely resemble the survey results.

The models from each type of network with the lowest score are shown in Figure 1[2]. The corresponding parameters are shown in Table 6.

*Figure 1: The final candidate models for each type of network with the lowest score value and the questionnaire results for the proportion of the binge drinker's friends who are thought to be binge drinkers*

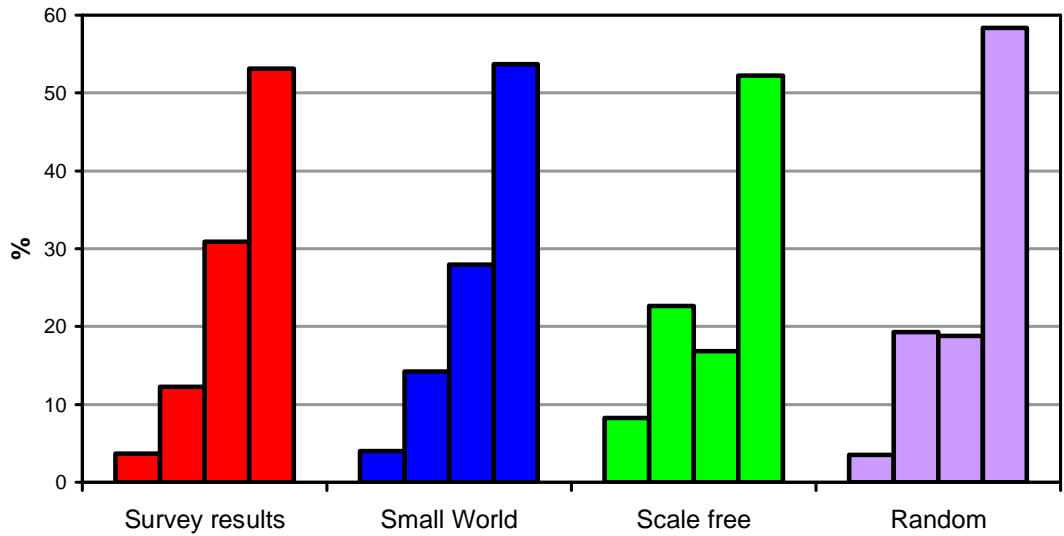

It can be seen that the candidate model for the small world most accurately reproduces the survey results and is therefore the chosen model. It significantly out performs the other types of networks including the random network which was used as a control. The candidate network models do not reproduce the profile of the quartiles.

---

[2] Note that only one type of small world network is shown, this is for the version with additional wiring. The level of rewiring is low in the small world $(\phi << 1)$ so results for both types of small world networks are the almost identical.

*Table 6: Parameters for the final candidate models*

| Network | Parameter | Optimised Value |
|---|---|---|
| Small world network | $k$ | 4 |
| | $\phi$ | 0.005 |
| | U1 | 0.5 |
| Scale free network | $q$ | 1 |
| | $\alpha$ | 4 |
| | U1 | 0.8 |
| Random network | $p$ | 0.002 |
| | U1 | 0.8 |

The optimized value of $k$ implies that in the context of drinking behaviour, binge drinkers regard 8 people as their friends.

The model rules explored so far have only considered social influence causing agents to take up binge drinking (the 0-1 transition), the possibility that social pressure could stimulate people to give up binge drinking (1-0 transitions) has not been explored.

In order to identify candidate models which had similar or better scores to the chosen model and that included 0-1 transitions the small world model space described in Table 6 was swept again with L2 values of 0.8, 0.6 and 0,5 and a value of U2 of 1.2 added to the combinations. This means that a proportion of agents will never be able to stop binge drinking if they take it up, irrespective of their network, but that the remaining fraction will give binge drinking if their network is sufficiently connected to non-binge drinkers.

The candidates for models including 1-0 transitions are shown in Figure 2 alongside the survey results and the chosen small world model. Two candidate models are shown, firstly a model with the same $k$, $\phi$ and U1 values as the chosen small world model and secondly the optimum model from the entire parameter space.

Figure 2 shows that the chosen model, without behavioural rules to give up binge drinking, outperforms the candidates of those that do. The fact that in both of the optimised candidate models the scale of social pressure to give up binge drinking is much lower that that to take it up provides more evidence of the robustness of the chosen model and that social influence to give up binge drinking (in this age group) can be approximated to zero.

*Figure 2: Results from introducing behavioural rules which allow agents to stop binge drinking based on their social network. Shown in red is the questionnaire results and blue is the optimised model. Green shows the model with the best score when U2 was introduced into the optimised model while the purple results show the best model from all available parameterisations.*

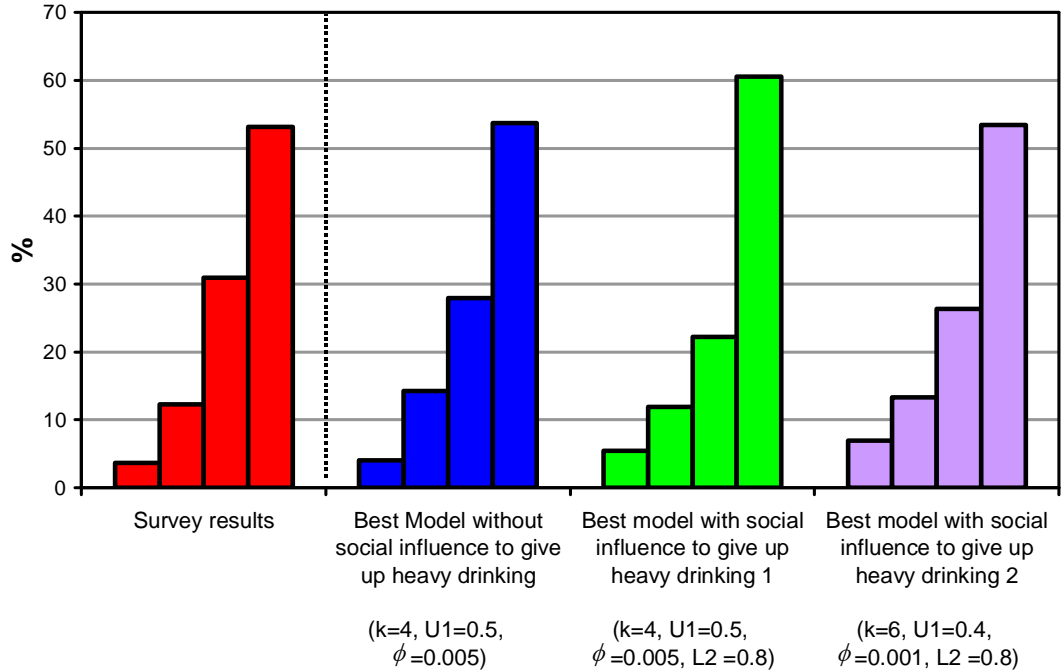

This result is intuitively plausible, given that the focus of the research is young people agents 18-24. The transition from 1 to 0 i.e. giving up binge drinking is likely to be closely connected to age, so as these particular agents get older, they will cease to be binge drinkers for a variety of reasons

## 5. Brief discussion

Binge drinking in the UK has grown rapidly in recent years and has become a matter of serious policy concern. 'Binge' means the rapid consumption of large amounts of alcohol, especially by young people, leading to anti-social behaviour in urban centres.

Increasingly, policy makers in the West are concerned about how not just to regulate but to alter social behaviour. Smoking and obesity are obvious examples, and in the UK 'binge' drinking has become a focus of acute policy concern.

We develop a simple agent based theoretical model which requires a limited amount of easily acquired information in order to calibrate scientifically. We examine whether the spread of imitative behaviour across friendship networks is a sufficient condition to account for the observed patterns of binge drinking behaviour in the UK.

A standard market research survey was carried out in order to discover both the number of binge drinkers in the 18-24 year old population, where the problem is most acute, and their friendship patterns in terms of drinking behaviour. There are decisive differences in the drinking behaviour of friends of binge drinkers compared to the drinking behaviour of non-binger drinkers.

We examined different types of potential networks, random, scale-free and small world with both additional and re-wired links. We conducted extensive searches for the best combination(s) of relevant parameters in each of the three types of networks considered.

A small world network was the optimal choice of network, and was able to generate a close approximation to the observed patterns of behaviour.

The research does not demonstrate that imitation on social networks is necessarily the only significant reason for the recent rapid and dramatic rise in binge drinking. But it offers strong evidence that this factor is important, indeed it is sufficient to describe current behaviour. So policy makers have to take this into account when they try to devise strategies to combat the problem.

The discovery that the relevant network has a small world structure is also helpful to policy makers. It does not tell them precisely what to do, but it suggests, for example, that strategies based upon the concept that there is a small number of 'influentials' who are important in the spread of this anti-social behaviour are not likely to be very successful.

If the network had been a scale-free one, then of course such an approach might well work very well, provided always that the 'influentials' can be identified. This finding provides empirical support for the theoretical proposition developed in [12] that that it is rarely the case that highly influential individuals are responsible for bringing about shifts in public opinion and/or behaviour.

**Acknowledgements**

We are grateful to the UK Advertising Association for sponsoring this research. The authors have full responsibility for the content.